\documentclass[aoas,nameyear,dvips]{arximspdf}

\doi{10.1214/09-AOAS312B}
\referstodoi{10.1214/09-AOAS312}
\volume{3}
\issue{4}
\pubyear{2009}
\firstpage{1270}
\lastpage{1278}

\makeatletter

\newtheorem{lemma}{Lemma}
\newproclaim{remark}{Remark}
\newproclaim{example}{Example}
\newcommand{\re}[0]{\mathbb{R}}
\newcommand{\parrow}[0]{\stackrel{P}{\rightarrow}}
\makeatother

\begin{document}
\begin{frontmatter}

\title{Discussion of: Brownian distance covariance\thanksref{T1}}
\pdftitle{Discussion on Brownian distance covariance by G. J. Szekely and M. L. Rizzo}
\thankstext{T1}{Supported in part by U.S. National
Institutes of Health Grant CA075142.}
\runtitle{Discussion}
\begin{aug}
\author{\fnms{Michael R.} \snm{Kosorok}\corref{}\ead[label=e1]{kosorok@unc.edu}}
\runauthor{M. R. Kosorok}
\affiliation{University of North Carolina at Chapel Hill}
\address{Department of Biostatistics and\\
\quad Department of Statistics and\\
\quad Operations Research \\
University of North Carolina at Chapel Hill\\
Chapel Hill, North Carolina 27599\\
USA\\
\printead{e1}} %adresu isvedimo komanda
%gale!
\end{aug}

% HISTORY:

% ABSTRACT
%
\begin{abstract}
We discuss briefly the very interesting concept of \mbox{Brownian} distance
covariance developed by Sz{\'{e}}kely and Rizzo [\textit{Ann. Appl.
Statist.} (2009), to appear] and describe two
possible extensions. The first extension is for high dimensional data
that can be coerced into a Hilbert space, including certain high
throughput screening and functional data settings. The second extension
involves very simple modifications that may yield increased power in
some settings. We commend Sz{\'{e}}kely and Rizzo for their very
interesting work and recognize that this general idea has potential to
have a large impact on the way in which statisticians evaluate
dependency in data.

\end{abstract}

% KEYWORDS
%
\begin{keyword}
\kwd{Brownian distance covariance}
\kwd{correlation}
\kwd{Hilbert spaces}
\kwd{$U$-statistics}.
\end{keyword}

\end{frontmatter}
%
%s1 ###
\section{Introduction and assessment}

The Brownian distance covariance and correlation proposed by Sz{\'
{e}}kely and Rizzo (\citeyear{Szekely09}) (abbreviated SR hereafter) is a very useful
and elegant alternative to the standard measures of correlation and is
based on several deep and nontrivial theoretical calculations developed
earlier in Sz{\'{e}}kely, Rizzo and Bakirov (\citeyear{Szekely07}) (abbreviated SRB
hereafter). We congratulate the group on this very original and elegant
work. The main result is that a single, simple statistic $\mathcal{V}_n(X,Y)$ can be used to assess whether two random vectors $X$ and
$Y$, of possibly different respective dimensions $p$ and $q$, are
dependent based on an i.i.d. sample.

The proposed statistic $\mathcal{V}_n(X,Y)$ estimates an interesting
population parameter $\mathcal{V}(X,Y)$ that the authors demonstrate can
also be expressed as the covariance between independent Brownian
motions $W$ and $W'$, with $p$ and $q$ dimensional indices, evaluated
at $X$ and $Y$, respectively. Specifically, let $W\dvtx \re^p\mapsto\re$ be
a real valued, tight, mean-zero Gaussian process with covariance
$|s|_p+|t|_p-|s-t|_p$, for $s,t\in\re^p$, where $|\cdot|_r$ is the
standard Euclidean norm in $\re^r$. Let $W'$ be similarly defined but
for indices $s,t\in\re^q$ and norm $|\cdot|_q$. It can be shown that
$\mathcal{V}(X,Y)=E[W(X)W(X')W'(Y)W'(Y')]$, where $(X',Y')$ is an
independent copy of $(X,Y)$, and where $W$ and $W'$ are independent of
both $(X,Y)$ and $(X',Y')$. This justifies the designation ``Brownian
distance covariance.''

By replacing Brownian motion with other stochastic processes, a very
wide array of alternative forms of correlation between vectors $X$ and
$Y$ can be generated. In the special case where $p=q=1$ and the
stochastic processes $W$ and $W'$ are the nonrandom identify functions
centered respectively at $E(X)$ and $E(Y)$, $\mathcal{V}_n(X,Y)=E[W(X)W(X')W'(Y)W'(Y')]=\operatorname{Cov}^2(X,Y)$, which is the
standard Pearson product-moment covariance squared. Thus, the results
obtained by SR not only have a profound connection to Brownian motion,
but also include traditional measures of dependence as special cases,
while, at the same time, having the potential to generate many useful
new measures of dependence through the use of other stochastic
processes besides Brownian motion. This raises the very real
possibility that a broadly applicable and unified theoretical and
methodological framework for testing dependence could be developed.

The SR paper is therefore not only important for the specific results
contained therein but also for the possibly far reaching consequences
for future statistical research in both theory and applications. For
the remainder of the paper, we describe two possible extensions of
these results. The first extension is for high dimensional data that
can be coerced into a Hilbert space, including certain high throughput
screening and functional data settings. The second extension involves
very simple modifications that may yield increased power in some
settings. We first present some initial results and consequences of SR
and SRB that will prove useful in later developments. We then present
the Hilbert space extension with a few example applications. Some
modifications leading to potential variations in power will then be
described. The paper will then conclude with a brief discussion.

%s2 ###
\section{Some initial results}
We now present a few initial results which will be useful in later
sections. For a paired sample of size $n$, $(X_1,Y_1),\ldots
,(X_n,Y_n)$, of realizations of $(X,Y)$, where $X$ and $Y$ are random
variables from arbitrary normed spaces with respective norms $\|\cdot\|
_X$ and $\|\cdot\|_Y$, define, analogously to SR,
\begin{eqnarray*}
T_1&=&\frac{1}{n^2}\sum_{k,l=1}^n\|X_k-X_l\|_X\|Y_k-Y_l\|_Y,\\
T_2&=&\frac{1}{n^2}\sum_{k,l=1}^n\|X_k-X_l\|_X\times\frac{1}{n^2}\sum
_{k,l=1}^n\|Y_k-Y_l\|_Y,\\
T_3&=&\frac{1}{n^3}\sum_{k=1}^n\sum_{l,m=1}^n\|X_k-X_l\|_X\|Y_k-Y_m\|_Y,
\end{eqnarray*}
and $V_n(X,Y)=T_1+T_2-2T_3$. Also define
\begin{eqnarray*}
T_{10}&=&E[\|X_1-X_2\|_X\|Y_1-Y_2\|_Y],\\
T_{20}&=&E[\|X_1-X_2\|_X]\times E[\|Y_1-Y_2\|_Y
],\\
T_{30}&=&E[\|X_1-X_2\|_X\|Y_1-Y_3\|_{Y}],
\end{eqnarray*}
and $V_0(X,Y)=T_{10}+T_{20}-2T_{30}$. Also let $V_n(X)=V_n(X,X)$ and
$V_0(X)=V_0(X,X)$; and let $V_n(Y)=V_n(Y,Y)$ and $V_0(Y)=V_0(Y,Y)$.
This allows us to define also $R_n(X,Y)=V_n(X,Y)/\sqrt{V_n(X)V_n(Y)}$
and $R_0(X,Y)=V_0(X,Y)/\break \sqrt{V_0(X)V_0(Y)}$, provided the denominators
are nonzero (and defined to be zero otherwise). The main distinction
between this and the definitions in SR is the use of arbitrary normed spaces.

Because this has a standard $U$-statistic structure, we have the
following general result, the proof of which follows from standard
theory for $U$-statistics [see, e.g., Chapter 12 of van der Vaart (\citeyear{Vaart98})]:
\begin{lemma}\label{lemma1}
Provided $E\|X\|_X^4<\infty$ and $E\|Y\|_Y^4<\infty$, then
$V_n(X,Y)\parrow V_0(X,Y)$, $V_n(X)\parrow V_0(X)$ and $V_n(Y)\parrow V_0(Y)$.
\end{lemma}

\begin{remark}\label{remark1}
In the special case where $X$ and $Y$ are from finite-dimensional
Euclidean spaces, we know from Theorems~1--4 of SR that $V_n(X,Y)$,
$V_n(X)$, $V_n(Y)$, $V_0(X,Y)$, $V_0(X)$ and $V_0(Y)$ are all
nonnegative; that $V_n(X,Y)\leq\break \sqrt{V_n(X)V_n(Y)}$ and $V_0(X,Y)\leq
\sqrt{V_0(X)V_0(Y)}$; that $V_0(X)=0$ or $V_0(Y)=0$ only when $X$ or
$Y$ is trivial; that $V_n(X)=0$ or $V_n(Y)=0$ only when the $X$'s or
$Y$'s in the sample are all identical; that $0\leq
R_n(X,Y),R_0(X,Y)\leq1$; and that $V_0(X,Y)=0$ only when $X$ and $Y$
are independent.
\end{remark}

We now wish to generalize the above results in the finite-dimensional
context to a class of norms more broad than Euclidean norms. These
results will be useful for later sections. Let $A$ and $B$ be
respectively $p\times p$ and $q\times q$ symmetric, positive definite
matrices. Let a ``tilde'' placed over $T_1$, $T_2$, $T_3$, $V_n$,
$V_0$, etc., denote the quantity obtained by replacing $|x|_p$ with $\|
x\|_{A,p}=\sqrt{x'Ax}$ and $|y|_q$ with $\|y\|_{B,q}=\sqrt{y'By}$ in
$V_n$, $V_0$, etc. For example,
$\tilde{T}_1=n^{-2}\sum_{k,l=1}^n\|X_k-X_l\|_{A,p}\|Y_k-Y_l\|_{B,q}$.
We now have the following very simple extension:
\begin{lemma}\label{lemma2}
Let $A$ and $B$ be symmetric and positive definite. Then $\tilde
{V}_n(X,Y)$, $\tilde{V}_n(X)$, $\tilde{V}_n(Y)$, $\tilde{V}_0(X,Y)$,
$\tilde{V}_0(X)$ and $\tilde{V}_0(Y)$ are all nonnegative; and all of
the other results in Remark~\ref{remark1} remain true with a ``tilde''
placed over the given quantities. Moreover, $\tilde{V}_0(X,Y)=0$ if and
only if $V_0(X,Y)=0$.
\end{lemma}

\begin{pf} For a symmetric, positive definite matrix $C$, let
$C^{1/2}$ denote the symmetric square root of $C$, that is,
$C^{1/2}C^{1/2}=C$. Note that such a square root always exists and,
moreover, is always positive definite. Now define $U=A^{1/2}X$ and
$V=B^{1/2}Y$, and note that $|U|_p=\|X\|_{A,p}$ and $|V|_q=\|Y\|
_{B,q}$. Now replace~$X$ and~$Y$ in the quantities listed in Remark~\ref
{remark1} with $U$ and $V$. By the symmetry properties of these norms,
the first part of the lemma up to just before the last sentence is
proved. The last sentence follows from the simple observation that $U$
and $V$ are independent if and only if $X$ and $Y$ are independent by
the positive definiteness of $A^{1/2}$ and $B^{1/2}$. Since
$V_0(X,Y)=0$ if and only if $X$ and $Y$ are independent, we now
conclude that $\tilde{V}_0(X,Y)=0$ if and only if $X$ and $Y$ are
independent. The entire lemma now follows.
\end{pf}

The third initial result involves some nontrivial properties of
independent components in the finite dimensional setting. Suppose for
$X\in\re^p$ and $Y\in\re^q$, where $p=p_1+p_2$ and $q=q_1+q_2$, we have
\[
X=\pmatrix{
X^{(1)}+X^{(2)}\cr X^{(3)}}  \quad \mbox{and} \quad Y=\pmatrix{
Y^{(1)}+Y^{(2)}\cr Y^{(3)}},
\]
where
$X^{(1)},X^{(2)}\in\re^{p_1}$, $X^{(3)}\in\re^{p_2}$,
$Y^{(1)},Y^{(2)}\in\re^{q_1}$, $y^{(3)}\in\re^{q_2}$; and suppose also
that the two vectors $\tilde{X}=([X^{(2)}]^T,[X^{(3)}]^T)^T$
and $\tilde{Y}=([Y^{(2)}]^T,[Y^{(3)}]^T)^T$ are mutually
independent and also independent of $X^{(1)}$ and $Y^{(1)}$. We have
the following somewhat surprising result:
\begin{lemma}\label{lemma3}
$V_0(X,Y)=V_0(X^{(1)},Y^{(1)})$.
\end{lemma}

\begin{pf} For any $t\in\re^p$ and $s\in\re^q$, with
$t=(t_1^T,t_2^T)^T$, $s=(s_1^T,s_2^T)^T$, $t_1\in\re^{p_1}$, $t_2\in\re
^{p_2}$, $s_1\in\re^{q_1}$, and $s_2\in\re^{q_2}$, the independence
assumptions and standard characteristic function properties yield
\begin{eqnarray*}
&&|E\exp(i[t^TX+s^TY])-E\exp(it^TX)E\exp(is^TY)|\\
&&\qquad=\bigl|f_{\tilde{X}}(t)f_{\tilde{Y}}(s)
\bigl\{E\exp\bigl(i\bigl[t_1^TX^{(1)}+s_1^TY^{(1)}\bigr]\bigr)\\
&&\qquad\quad\hspace*{57pt} {}-E\exp\bigl(it_1^TX^{(1)}\bigr)E\exp
\bigl(is_1^TY^{(1)}\bigr)\bigr\}\bigr|\\
&&\qquad=\bigl|E\exp\bigl(i\bigl[t_1^TX^{(1)}+s_1^TY^{(1)}\bigr]\bigr)-E\exp\bigl(it_1^TX^{(1)}\bigr)E\exp
\bigl(is_1^TY^{(1)}\bigr)\bigr|\\
&&\qquad=
|f_{X^{(1)},Y^{(1)}}(t_1,s_1)-f_{X^{(1)}}(t_1)f_{Y^{(1)}}(s_1)|.
\end{eqnarray*}
Combining this with Theorems~1 and~2 of SR, we obtain that
\[
V_0(X,Y)=\frac{1}{c_pc_q}\int_{\re^{p+q}}\frac{
|f_{X^{(1)},Y^{(1)}}(t_1,s_1)-f_{X^{(1)}}(t_1)f_{Y^{(1)}}(s_1)
|^2}{|t|_p^{p+1}|s|_q^{q+1}}\,dt\,ds.
\]
Note that the right-hand side is invariant with respect to the
distributions of $\tilde{X}$ and~$\tilde{Y}$ and, thus, we can replace
$\tilde{X}$ and~$\tilde{Y}$ with degenerate random variables fixed at
zero. Doing the same on the left-hand side yields the desired
result.
\end{pf}

%s3 ###
\section{High dimensional extensions}

The basic idea we propose is to extend the results to Hilbert spaces
which can be approximated by sequences of finite-dimensional Euclidean
spaces. We will give a few examples shortly. First, we give the
conditions for our results. Assume $X$ is a random variable in a
Hilbert space~$H_X$ with inner produce $\langle\cdot,\cdot\rangle_X$
and norm $\|\cdot\|_X$. A superscript $\ast$ will be used to denote
adjoint. Say that $X$ is ``finitely approximable'' if there exists a
sequence $X_m\in H_X$ such that for each $m\geq1$, there exists a
linear map $M_m\dvtx H_x\mapsto\re^{p_m}$ for which $M_m^{\ast}M_m$ is
symmetric and positive definite on $\re^{p_m}$, $p_m$ is nondecreasing,
$X_m=M_m(U_m)$ for some sequence of Euclidean random variables $U_m$,
and that $E\|X_m-X\|_X^2\rightarrow0$ as $m\rightarrow\infty$. Note
that we can assume that $M_m^{\ast}M_m$ is the identity without loss of
generality. This follows since we can always replace $U_m$ with $\tilde
{U}_m=A_mU_m$ and $M_m$ with $\tilde{M}_m=M_mA_m^{-1}$, where
$A_m=(M_m^{\ast}M_m)^{1/2}$, to yield $X_m=\tilde{M}_m\tilde{U}_m$ with
$\tilde{M}_m^{\ast}\tilde{M}_m=A_m^{-1}(M_m^{\ast}M_m)A_m^{-1}$ being
the identity.

\begin{example}\label{ex1} Let $X$ be functional data with realizations that are
functions in the Hilbert space $H_X=L_2[0,1]$ consisting of functions
$f\dvtx [0,1]\mapsto\re$ satisfying $\|f\|_X^2=\int_0^1 f^2(t)\,dt<\infty$.
Specifically, we will assume that
%
%e1 ###
\begin{equation}
X(t)=\sum_{i=1}^{\infty}\lambda_iZ_i\phi_i(t),\label{eq1}
\end{equation}
where $Z_1,Z_2,\ldots$ are independent random variables with mean zero
and variance~1, $\phi_1,\phi_2,\ldots$ form an orthonormal basis in
$L_2[0,1]$, and $\lambda_1,\lambda_2,\ldots$ are fixed constants
satisfying $\sum_{i=1}^n\lambda_i^2<\infty$. This formulation can yield
a large variety of tight stochastic processes and can be a realistic
model for some kinds of functional data.

Let $p_m=m$, $U_m=(\lambda_1Z_1,\ldots,\lambda_mZ_m)^T$, and, for any
vector $a\in\re^{p_m}$, $M_m(a)=\sum_{i=1}^ma_i\phi_i(t)$.
Clearly, $X_m=M_m(U_m)$ is in $H_X$ almost surely, since $\|X_m\|_X=\sum
_{i=1}^m\lambda_i^2Z_i^2$ is bounded almost surely. Moreover, for any
$f\in L_2[0,1]$, it can be shown that
\[
M_m^{\ast}(f)=\pmatrix{
\int_0^1\phi_1(s)f(s)\,ds\cr
\vdots\cr
\int_0^1\phi_m(s)f(s)\,ds},
\]
and, thus, $M_m^{\ast}M_m$ is the identity by the orthonormality of the
basis and is therefore positive definite. Since $\sum_{i=1}^{\infty
}\lambda_i^2<\infty$,
\begin{eqnarray*}
E\|X-X_m\|_X^2&=&E\Biggl\|\sum_{i=m+1}^{\infty}\lambda_iZ_i\phi
_i(t)\Biggr\|_X^2\\
&=&\sum_{i=m+1}^{\infty}\lambda_i^2\\
&\rightarrow& 0,
\end{eqnarray*}
as $m\rightarrow\infty$. Thus, $X$ is finitely approximable.
\end{example}

\begin{example}\label{ex2} This is basically the same as Example~\ref{ex1}, except that
we will not require the basis functions to be orthogonal. Specifically,
let $X(t)$ be as given in~(\ref{eq1}), with the basis functions
satisfying $\int_0^1 \phi_i^2(s)\,ds=1$, for all $i\geq1$, but not
necessary being mutually orthogonal. Let $a_{i,j}=\int_0^1\phi_i(s)\phi
_j(s)\,ds$, for $i,j\geq1$, and define $A_m$ to be the $m\times m$
matrix with entry $a_{i,j}$ for row $i$ and column $j$ for $1\leq
i,j\leq m$. Assume that $A_M$ is positive definite for each $m\geq1$ and
also assume that $\lim_{m\rightarrow\infty}\sum_{i,j=m+1}^{\infty
}\lambda_i\lambda_ja_{i,j}=0$.
If we now follow parallel calculations to those done in Example 1, we
can readily deduce that with $X_m=\sum_{i=1}^m\lambda_iZ_i\phi_i(t)$,
we have $M_m$ and $M_m^{\ast}$ defined as before, but with $M_m^{\ast
}M_m=A_m$ instead of the identity, while $E\|X-X_m\|_X^2\rightarrow0$
also as before. The increased flexibility enlarges the scope of
stochastic processes achievable to include, for example, Brownian motion.
\end{example}

\begin{example} Let $X=(X^{(1)},X^{(2)},\ldots)^T$ be an
infinitely long Euclidean vector in $\ell_2$, that is, $\sum
_{i=1}^{\infty}[X^{(i)}]^2<\infty$ almost surely; and assume
that, after permuting the indices if necessary,
\[
\sum_{i=m+1}^{\infty}E\bigl[X^{(i)}\bigr]^2\rightarrow0,
\]
as $m\rightarrow\infty$. It is fairly easy to see that if we let $X_m$
be a vector with the first $m$ elements being identical to the first
$m$ elements of $X$ but with all remaining elements equal to zero, then
$E\|X-X_m\|_X^2\rightarrow0$, as $m\rightarrow\infty$, and all of the
remaining conditions for finite approximability are satisfied. This
example may be applicable to certain high throughput screening settings
where the vector of measurements may be arbitrarily high-dimensional.
\end{example}

The following lemma tells us that the range-related properties of
Brownian distance covariance are preserved for finitely approximable
random variables:
\begin{lemma}\label{lemma4}
Assume that $X$ and $Y$ are both finitely approximable random variables
in Hilbert spaces. Then $V_n(X,Y)$, $V_n(X)$, $V_n(Y)$, $V_0(X,Y)$,
$V_0(X)$ and $V_0(Y)$ are all nonnegative, $V_n(X,Y)\leq\sqrt
{V_n(X)V_n(Y)}$, $V_0(X,Y)\leq\break\sqrt{V_0(X)V_0(Y)}$, and $0\leq
R_n(X,Y),R_0(X,Y)\leq1$.
\end{lemma}

\begin{pf} Let $X_m$ and $Y_m$ be sequences such that $E\|X-X_m\|
_X^2\rightarrow0$ and $E\|Y-Y_m\|_Y^2\rightarrow0$ as $m\rightarrow
\infty$. Using simple algebra, we can verify that
$V_0(X_m,Y_m)\rightarrow V_0(X,Y)$ which implies $V_0(X,Y)\geq0$.
Similar arguments verify the desired results for $V_0(X)$, $V_0(Y)$ and
$R_0(X,Y)$. Now, for a sample of size~$n$, $(X_1,Y_1),\ldots
,(X_n,Y_n)$, we can create a sequence of samples $(X_{1m},Y_{1m}),\ldots
, (X_{nm},Y_{nm})$, such that $\sum_{i=1}^n(E\|X_i-X_{im}\|_X^2+E\|
Y_i-Y_{im}\|_Y^2)\rightarrow0$ by finite approximability. Let
$V_n^{(m)}(X,Y)$ be the same as $V_n(X,Y)$ but with the $m$th
approximating sample replacing the sample observations. Since
convergence in mean implies convergence in probability, we can apply
basic algebra to verify that $V_n^{(m)}(X,Y)\parrow V_n(X,Y)$ as
$m\rightarrow\infty$. Similar arguments verify the desired results for
$V_n(X)$, $V_n(Y)$ and $R_n(X,Y)$, and this completes the proof.
\end{pf}

Our ultimate goal in this section, however, is to show that $R_0(X,Y)$
has the same implications for assessing dependence for finitely
approximable Hilbert spaces as it does for finite dimensional settings.
This is actually quite challenging, and we are only able to achieve
part of the goal in this paper. The following is our first result in
this direction:
\begin{lemma}\label{lemma5}
Suppose $X$ and $Y$ are random variables in finitely approximable
Hilbert spaces. Then $R_0(X,Y)>0$ implies that $X$ and $Y$ are dependent.
\end{lemma}

\begin{pf} Assume that $R_0(X,Y)>0$ but that $X$ and $Y$ are
independent. By finite approximability, there exists a sequence of
paired random variables $(X_m,Y_m)$ such that $X_m$ and $Y_m$ are
independent for each $m\geq0$, $E\|X-X_m\|_X^2\rightarrow0$, and $E\|
Y-Y_m\|_Y^2\rightarrow0$. This implies that $R_0(X_m,Y_m)=0$ for all
$m\geq0$. Since also $R_0(X_m,Y_m)\rightarrow R_0(X,Y)$, we have a
contradiction. Hence, $X$ and $Y$ are dependent.
\end{pf}

If we could also show that $R_0(X,Y)=0$ implies independence, we would
have essentially full homology with the finite dimensional case. It is
unclear how to show this in general, and it may not even be true in
general. However, it is certainly true for an interesting special case
which we now present.

Let $X$ and $Y$ be random variables in finitely approximable Hilbert
spaces. Suppose there exists linear maps $M\dvtx H_X\mapsto H_X$ and
$N\dvtx H_Y\mapsto H_Y$ with adjoints for which both $M^{\ast}M$ and $N^{\ast
}N$ are identities, and that $MX=X_1+X_2$ and $NY=Y_1+Y_2$, where
$X_1\in H_X^{(1)}$ and $Y_1\in H_Y^{(1)}$, $H_X^{(1)}$ and $H_Y^{(2)}$
are finite-dimensional subspaces of $H_X$ and $H_Y$, respectively, and
that $X_2$ and $Y_2$ are mutually independent and independent of
$(X_1,Y_1)$. We will call a random pair $(X,Y)$ that satisfies these
conditions ``at most finitely dependent.'' For example, paired
functional data $(X,Y)$ could be at most finitely dependent if all
possible dependencies between the two populations $X$ and $Y$ are
attributable to at most a few principle functions (or principle
components) in each population and that the remaining components are
independent noise.

\begin{example} Suppose that we are interested in determining whether
$X$ and $Y$ are independent, where $X$ is either a functional
observation or some other very high dimensional observation and $Y$ is
a continuous outcome of interest such as a time to an event. Suppose
also that $X$ is finitely approximable and that any potential
dependence of $Y$ on $X$ is solely due to a latent set of finite
principle components of $X$. Such a pair $(X,Y)$ would be at most
finitely dependent.
\end{example}

The following lemma on finitely dependent data is the final result of
this section:
\begin{lemma}\label{lemma6}
Suppose that $X$ and $Y$ are finitely approximable random variables in
Hilbert spaces and that $(X,Y)$ is at most finitely dependent. Then
$R_0(X,Y)\geq0$ and the inequality is strict if and only if $X$ and
$Y$ are dependent.
\end{lemma}

\begin{pf} Note first that $\|MX\|_X^2=\langle MX, MX\rangle
_X=\langle M^{\ast}MX, X\rangle_X=\langle X,\break  X\rangle_X=\|X\|_X^2$ and,
similarly, $\|NY\|_Y=\|Y\|_Y$. Since $R_0(X,Y)$ is a function involving
only the norms of $X$ and $Y$, we can assume without loss of generality
that $N$ and $M$ are identities. Thus, we will simply assume that
$X=X_1+X_2$ and $Y=Y_1+Y_2$ hereafter. Let $(X_{2m},Y_{2m})$ be a
sequence of paired random variables in $H_X\times H_Y$ such that $E\|
X_2-X_{2m}\|_X^2\rightarrow0$ and $E\|Y_2-Y_{2m}\|_Y^2\rightarrow0$,
and where, for each $m\geq1$, $X_{2m}$ and $Y_{2m}$ are mutually
independent and also independent of $(X_1,Y_1)$.

Now let $\hat{X}_m=X_1+X_{2m}$ and $\hat{Y}_m=Y_1+Y_{2m}$, and note
that both $\hat{X}_m$ and $\hat{Y}_m$ are finite dimensional with
$R_0(\hat{X}_m,\hat{Y}_m)\rightarrow R_0(X,Y)$. Let $p_1$ and $q_1$ be
the respective dimensions of $X_1$ and $Y_1$, $p_{2m}$ and $q_{2m}$ be
the respective dimensions of $X_{2m}$ and~$Y_{2m}$, and let
$p_m=p_1+p_{2m}$ and $q_m=q_1+q_{2m}$.
Let $X_{2m}^{(1)}$ be the projection of $X_{2m}$ onto $H_X^{(1)}$,
$Y_{2m}^{(1)}$ be the projection of~$Y_{2m}$ onto $H_{Y}^{(1)}$, and
let $X_{2m}^{(2)}=X_{2m}-X_{2m}^{(1)}$ and
$Y_{2m}^{(2)}=Y_{2m}-Y_{2m}^{(1)}$. By the finite-dimensionality of
$X_1$, $X_{2m}$, $Y_1$ and $Y_{2m}$, there exists linear maps $A_1\dvtx \re
^{p_1}\mapsto H_X^{(1)}$, $A_{2m}\dvtx \re^{p_{2m}}\mapsto H_X$,
$B_1\dvtx \re^{q_1}\mapsto H_Y^{(1)}$, and $B_{2m}\dvtx \re^{q_{2m}}\mapsto H_Y$,
such that $A_1^{\ast}A_1$, $A_{2m}^{\ast}A_{2m}$, $B_1^{\ast}B_1$ and
$B_{2m}^{\ast}B_{2m}$ are all identities and that $X_1=A_1U_1$,
$X_{2m}^{(1)}=A_1U_{2m}^{(1)}$, $X_{2m}^{(2)}=A_{2m}U_{2m}^{(2)}$,
$Y_1=B_1Z_1$, $Y_{2m}^{(1)}=B_1Z_{2m}^{(1)}$,
and $Y_{2m}^{(2)}=B_{2m}Z_{2m}^{(2)}$, for random vectors
$U_1,U_{2m}^{(1)}\in\re^{p_1}$, $U_{2m}^{(2)}\in\re^{p_{2m}}$,
$Z_1,Z_{2m}^{(1)}\in\re^{q_1}$, and $Z_{2m}^{(2)}\in\re^{q_{2m}}$,
where $U_{2m}=([U_{2m}^{(1)}]^T,[U_{2m}^{(2)}]^T)^T$ and
$Z_{2m}=([Z_{2m}^{(1)}]^T,[Z_{2m}^{(2)}]^T)^T$ are mutually
independent and independent of $(U_1,Z_1)$.

If we let $\hat{U}_m=([U_1+U_{2m}^{(1)}]^T,[U_{2m}^{(2)}]^T
)^T$ and $\hat{Z}_m=([Z_1+Z_{2m}^{(1)}]^T,[Z_{2m}^{(2)}]^T
)^T$, the above formulation yields that $\|\hat{X}_m\|_X=|\hat
{U}_m|_{p_m}$ and $\|\hat{Y}_m\|_Y=|\hat{Z}_m|_{q_m}$. By Lemma~\ref
{lemma3}, we now have that $R_0(\hat{U}_m,\hat{Z}_m)=R_0(U_1,Z_1)$
which does not depend on $m$. Since $A_1^{\ast}A_1$ and $B_1^{\ast}B_1$
are both identities, we also have that $R_0(U_1,Z_1)=R_0(X_1,Y_1)$ and,
thus, $R_0(\hat{X}_m,\hat{Y}_m)=R_0(\hat{U}_m,\hat{Z}_m)\rightarrow
R_0(X_1,Y_1)$, as $m\rightarrow\infty$. This now implies that
$R_0(X,Y)=R_0(X_1,Y_1)$, which yields the desired result. %\rightqed
\end{pf}

%s4 ###
\section{Increasing power}

We now briefly discuss the issue of power of tests based on $R_n(X,Y)$.
By Lemma~\ref{lemma2}, we observe that there are many different
versions of the statistic $R_n(X,Y)$, based on different choices of
matrices $A$ and $B$ in the norms $\|\cdot\|_{A,p}$ and $\|\cdot\|
_{B,q}$, that all have the ability to assess general dependence. Is it
possible to choose $A$ and $B$ in a way that provides optimal power for
certain fixed or contiguous alternatives? The answer should be yes
since it appears that $A$ and $B$ could potentially be selected to
emphasize dependence for certain subcomponents of $X$ and $Y$ while
de-emphasizing dependence for other subcomponents. The answer to this
question, unfortunately, seems to be very hard to pin down rigorously.
We do not pursue this further here, but it does seem to be a
potentially important issue that deserves further attention.

%s5 ###
\section{Discussion}

We have briefly proposed two generalizations of the Brownian distance
covariance, one based on alternative norms to Euclidean norms, and the
other based on infinite dimensional data. The first generalization
raises the possibility of fine-tuning the statistics proposed in SR to
increase power, and the second generalization opens the door for
applicability of the results in SR to a broader array of data types,
including infinite dimensional data and data with dimension increasing
with sample size. However, for both of these generalizations, there
remain many open questions that could lead to important further
improvements. In either case, the results of SR are very important both
practically and theoretically and should result in many important
future developments in both the application and theory of statistics.

\printaddresses

\end{document}

\begin{frontmatter}

\title{Correction: Discussion of
Brownian distance covariance}

\pdftitle{Correction: Discussion of Brownian distance covariance}
\runtitle{Correction}

\begin{aug}
\author[A]{\fnms{Michael R.} \snm{Kosorok}\corref{}\ead[label=e1]{kosorok@unc.edu}}
\runauthor{M.~R. Kosorok}
\affiliation{University of North Carolina at Chapel Hill}
\address[A]{Department of Biostatistics\\
University of North Carolina at Chapel Hill\\
Chapel Hill, North Carolina 27599\\
USA\\
\printead{e1}} %adresu isvedimo komanda gale!
\end{aug}

% HISTORY:
\received{\smonth{2} \syear{2013}}

% ABSTRACT

% KEYWORDS

\end{frontmatter}
The proof of Lemma~3 in \citet{Kosorok09} is incorrect since the second
equality in the display of the proof is, in fact, an inequality ($\leq
$). Some results in \citet{Dueck12} also indicate that Lemma~3 is not
true. Moreover, it is not hard to obtain simple counterexamples. For
example, if $X^{(2)}$ is a Rademacher random variable [i.e.,
$P(X^{(2)}=-1)=P(X^{(2)}=1)=1/2$] and $X^{(3)}$ is zero with
probability 1, then $f_{\tilde{X}}(t)=\cos(t)$ and the inequality is
strict for all $t$ for which $|\cos(t)|\neq1$.

Nevertheless, the conclusions of Lemma~6 in \citet{Kosorok09} remain
valid, as shown in \citet{LyonsR}, under even weaker conditions than
those given in the statement of the lemma. Moreover, the other results
of \citet{Kosorok09} are unaffected by Lemma~3 and thus remain valid.

\section*{Acknowledgments}
Thanks to Steven Hoberman, Dominic Edelmann and Tilmann Gneiting for
drawing my
attention to this error. Thanks also to Russell Lyons for making me
aware of his results.

%
% imsref loaded by akundreckaite, 2013-04-11 13:57:31


%suskaldyti doi

\printaddresses